\documentclass[11pt]{scrartcl}
\usepackage[latin1]{inputenc}
\usepackage[dvips]{graphicx,epsfig,color}
\usepackage{wrapfig,rotating}
\usepackage{amssymb,amsmath,array}
\usepackage{cite}
\begin{document}
\title{
%%%%   Paper title goes here  %%%%%%%%%%%%%%
%
R\&D Status of a gas-compressor based two-phase
CO$_2$ cooling system for FPCCD Vertex Detector }
%***********************************************************************
% AUTHORS INFORMATION AREA
%***********************************************************************
\author{Yasuhiro Sugimoto$^1$\footnote{Talk presented at the 
International Workshop on Future Linear Colliders (LCWS2016),
Morioka, Japan, 5--9 December 2016. C16-12-05.4}, 
Keisuke Fujii$^1$, 
Takahiro Fusayasu$^2$,\\
Katsuyu Kasami$^1$, and Tohru Tsuboyama$^1$ 
% Optional short acknowledgment: remove next line if non-needed
%\thanks{This is an optional funding source acknowledgment.}
% DO NOT MODIFY THE FOLLOWING '\vspace' ARGUMENT
\vspace{.3cm}\\
% Addresses and institutions (remove "1- " in case of a single institution)
1- High Energy Accelerator Research Organization (KEK) \\
Tsukuba, Ibaraki 305-0801, Japan
%% Remove the next three lines in case of a single institution
\vspace{.1cm}\\
2- Department of Physics, Saga University \\
Saga, 840-8502, Japan
\date{}
}
%%***********************************************************************
% END OF AUTHORS INFORMATION AREA
%***********************************************************************
%
\maketitle
\begin{abstract}
Fine pixel CCD (FPCCD) is one of the candidate sensor
technologies for the vertex detector used for experiments
at the International Linear Collider (ILC).
FPCCD vertex detector is supposed to be cooled
down to $-40^\circ$C for improvement of radiation immunity. 
For this purpose, a two-phase CO$_2$ cooling system
using a gas compressor for CO$_2$ circulation is
being developed at KEK. The status of this R\&D is 
presented in this article. 
\end{abstract}
\section{Introduction}
There are many sensor technologies proposed for the use
in the vertex detector for experiments at the International
Linear Collider (ILC)~\cite{behnke}. 
Fine pixel CCD (FPCCD) is one of the
candidates~\cite{sugimoto05}. 
Due to its small pixel size of $\sim 5\ \mu$m, 
FPCCD vertex detector can offer very good 
impact parameter resolution and excellent two-track separation
capability. FPCCD sensors will be read out 
in $\sim 200$~ms between bunch trains.
Because there is no beam crossing during the readout,
FPCCD sensor option is completely free from the RF noise
caused by the very short bunches of the beam.
One drawback of FPCCD sensors is relatively
poor radiation immunity, particularly large charge transfer
inefficiency (CTI) due to radiation induced trap levels.

\begin{figure}
\centerline{\includegraphics[width=0.55\columnwidth]{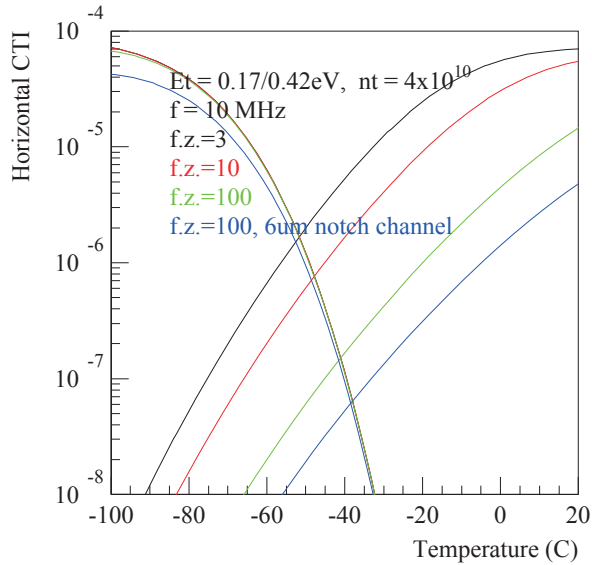}}
\caption{Results of a simple simulation of CTI with trap levels of 0.17~eV
and 0.42~eV based on Shockley-Read-Hall theory.
With fat-zero charge injection of 100 electrons and notch channel, 
$\sim -40^\circ$C
is the optimum operation temperature.}
\label{fig:hcti}
%\end{wrapfigure}
\end{figure}

CTI of FPCCD due to radiation damage is a function of temperature.
A simple simulation of CTI based on Shockley-Read-Hall theory
shows that around $-40^\circ$C is the optimal operating temperature
as shown in Figure~\ref{fig:hcti}.
To achieve this cooling temperature, two-phase CO$_2$ cooling system 
seems most suited.
%\begin{wrapfigure}{r}{0.45\columnwidth}

\section{Advantages of two-phase CO$_2$ cooling system}
In cooling systems using two-phase coolant,
the cooling temperature is controlled by controlling
the pressure of the two-phase coolant. 
Because the heat load is consumed only for evaporation of the
coolant, the cooling temperature is constant along
the cooling tube.
Perfluorocarbon (PFC) such as C$_2$F$_6$ has been 
also used as two-phase coolant for detector cooling systems,
for example ATLAS SCT~\cite{attree}. 
Compared with PFC, CO$_2$ has several 
advantages.
Table~\ref{tab:coolant} shows properties of CO$_2$ and
PFC. 
Latent heat of CO$_2$ is much larger than that of PFC.
Because the pressure of two-phase CO$_2$ is higher than
that of PFC at the same temperature, 
CO$_2$ has less temperature drop due to 
pressure drop along the cooling tube, and less
evaporated  gas volume than those of PFC. Thanks to
these properties, we can use thinner cooling tube 
for two-phase CO$_2$ than PFC. Outer diameter of 2~mm
or less is good enough for the cooling tube of FPCCD
vertex detector. 

Because heat sources of FPCCD vertex detector
locate mainly at both ends of ladders (on-chip amplifiers and 
front-end ASICs), cooling at both ends of ladders 
through the ladder base made of carbon fiber reinforced plastic
(CFRP) sheet and endplates on which cooling tube is attached
is an attractive solution.
The support shell including the end-plate is enclosed in a
cryostat made of heat insulating material.  
The return line of CO$_2$ will be used to cool the electronics
(clock drivers and data compression circuits) placed 
outside the cryostat.
The power consumption of the FPCCD vertex detector will be
less than 100~W inside the cryostat and about 200~W/side 
for the electronics outside the cryostat.
Additional material budget
due to attached cooling tube of 2~mm$\phi$ made of
Titanium would be only 0.3\%$X_0$ if averaged over 
the end-plate.

Gas cooling is a possible alternative. 
However, if we try to cool this vertex detector with
cold air or nitrogen gas, the flow rate will be 
quite large. As a consequence, vibration of 
the ladders could be caused. 
Much thicker transfer tube than two-phase CO$_2$ cooling
is necessary for gas cooling,
which  requires more dead space between forward tracking discs (FTD) 
and the beam pipe.
In addition, constant temperature cooling is almost impossible
in case of gas cooling.
\begin{table}
\centerline{\begin{tabular}{|l|l|l|l|}
\hline
         & CO$_2$ & C$_2$F$_6$ & C$_3$F$_8$ \\
\hline 
Latent heat @$-40^\circ$C [J/g]  & 321  & $\sim 100$ & $\sim 110$ \\ 
Triple point [$^\circ$C]  & $-56.4$ & $-97.2$ & $-160$  \\ 
Critical point [$^\circ$C] & 31.1 & 19.7 & 71.9 \\ 
Pressure @$-40^\circ$C [MPa] & 1 & $\sim 0.5$ & $\sim 0.1$ \\
Global warming potential & 1 & 12200 & 8830\\ 
\hline
\end{tabular}}
\caption{Properties of several kinds of two-phase coolant.}
\label{tab:coolant}
\end{table}
\section{Gas-compressor based cooling system}
Detector cooling systems using liquid pump for
circulation of  two-phase CO$_2$ have been
developed by several groups~\cite{delli,beuzekom}. 
In such a 
``pump-based system'', CO$_2$ is liquefied 
at temperature below the cooling temperature.
The temperature of circulating CO$_2$ is
below or at cooling temperature.
Therefore, very tight thermal insulation 
is required for the whole system, including
the transfer tubes and the liquid pump.
An expensive low-temperature chiller is 
necessary for liquefaction if the cooling temperature
as low as $-40^\circ$C has to be achieved.
A sophisticated ``two-phase accumulator''
has to be adopted for pressure (and temperature)
control of the two-phase CO$_2$ for the 
pump-based system.

\begin{figure}
\centerline{\includegraphics[width=0.75\columnwidth]{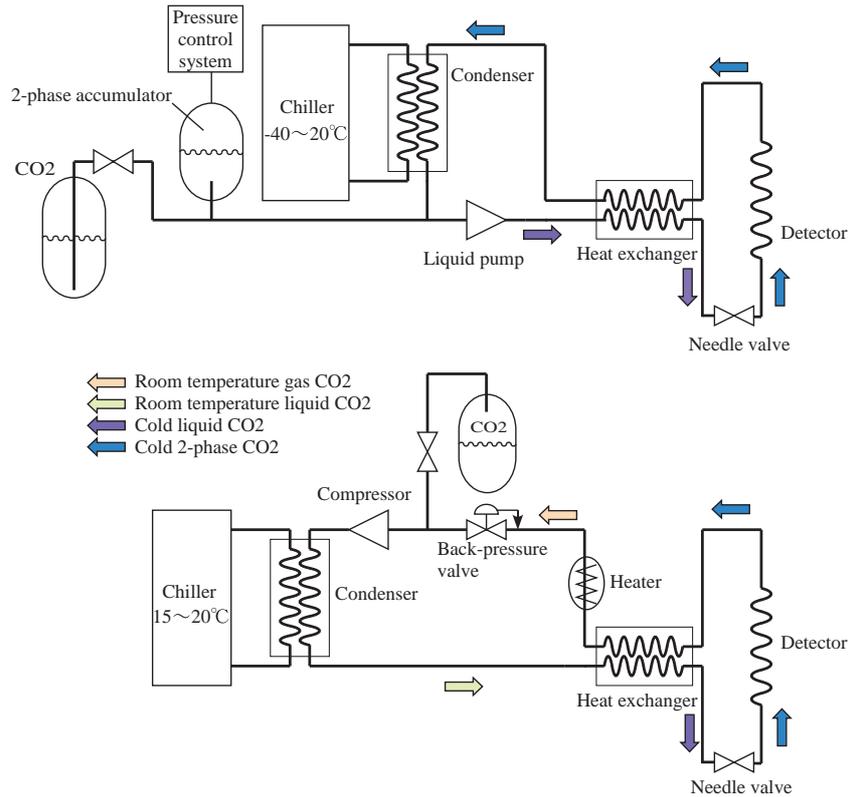}}
\caption{Schematic diagrams of a pump-based cooling system (top)
and a gas-compressor based cooling system (bottom).}
\label{fig:schema}
%\end{wrapfigure}
\end{figure}

Our R\&D goal is to develop a two-phase CO$_2$ 
cooling system using a gas compressor,
rather than a liquid pump,  for
circulation of CO$_2$. 
Figure~\ref{fig:schema} shows
schematic diagrams of a pump-based system and 
a gas-compressor based system.
In the gas-compressor based system, 
CO$_2$ gas is liquefied by a condenser at near room temperature
after compression.
The liquid CO$_2$ is transferred to a
heat exchanger near the detector through a
transfer tube.   
The long transfer tube between the liquefier plant
and the heat exchanger can be at near room temperature.
At the heat exchanger, the liquid CO$_2$ is cooled down
to the detector cooling temperature by the 
returning two-phase CO$_2$. 
Then, the pressure of the CO$_2$ is decreased 
by a needle valve (or a capillary tube).
The cooling is achieved basically by the latent heat 
(evaporation) of the returning 
two-phase CO$_2$,
rather than the Joule-Thomson effect.
The  two-phase CO$_2$ is completely evaporated
by a heater, and goes back to the liquefier plant.
The pressure of the two-phase CO$_2$ is controlled
by a back pressure valve in the liquefier plant.

The gas-compressor system has several advantages
over the pump system. Because CO$_2$ is liquefied
at near room temperature, we don't need an expensive  
low temperature chiller. Cooling water near room temperature
is enough for the liquefaction. In case of ILC experiment,
such cooling water must be available in the detector hall.
We don't need strict thermal insulation for long transfer tubes
between the liquefier plant and the detector. 
Flexible transfer tubes 
off-the-shelf can be used for this purpose. 
These tubes can be placed on the cable chain
supposed to be used for push-pull operation of 
ILC detectors~\cite{behnke}.
%\begin{wrapfigure}{r}{0.45\columnwidth}

\section{Construction of a prototype}
\begin{figure}
\centerline{\includegraphics[width=0.85\columnwidth]{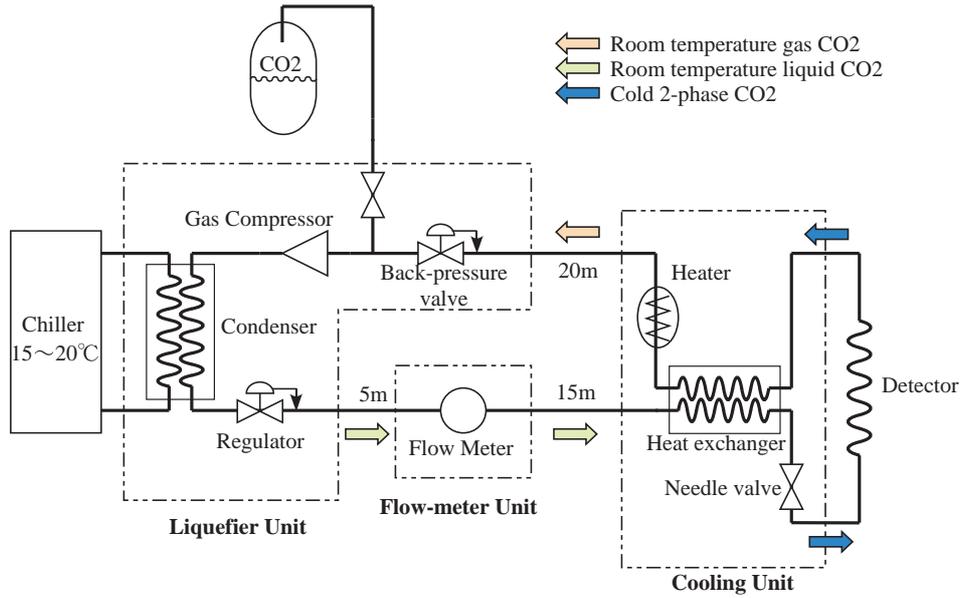}}
\caption{Schematic diagrams of the prototype cooling system.}
\label{fig:prototype}
%\end{wrapfigure}
\end{figure}

Several prototypes of the two-phase CO$_2$ cooling system
have been constructed at KEK~\cite{sugimoto12,sugimoto12b,sugimoto13}.
The latest prototype consists of three units;
a liquefier unit, a flow meter unit, and a cooling unit.
Figure~\ref{fig:prototype} shows
a simplified schematic diagram of the prototype system.
Three units are connected with metal-core flexible tubes
($1/4$ inch for liquid, $3/8$ inch for gas).
As a gas compressor, Haskel gas booster AGD-7 is used.
This compressor is a reciprocating type, and driven
by compressed air. The exhausted air is used for
cooling of gas cylinder of the compressor. 
%\begin{wrapfigure}{r}{0.45\columnwidth}

\begin{figure}
\centerline{\includegraphics[width=0.98\columnwidth]{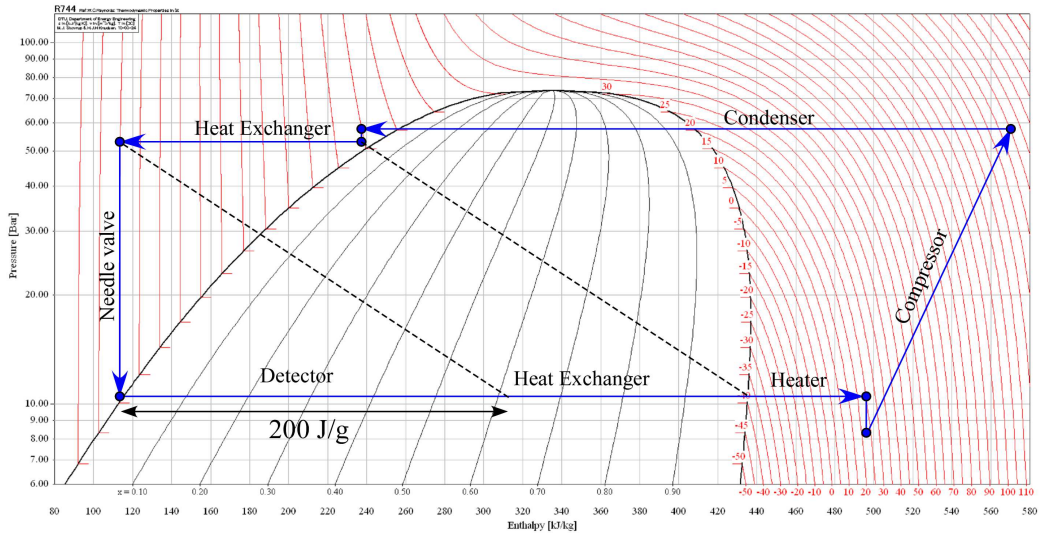}}
\caption{Phase diagram (p-H curve) of the cooling system in an ideal operation
condition.}
\label{fig:phasediagram}
%\end{wrapfigure}
\end{figure}

The phase diagram of the system in an ideal operating 
condition is shown in 
Figure~\ref{fig:phasediagram}.
The expected cooling power at $-40^\circ$C is 200~J/g.
The actual cooling power has been measured using dummy load
by looking at the dry-out point where the temperature of
CO$_2$ starts increasing.
The two-phase CO$_2$ with  flow rate of 1.4~g/s
at cooling temperature of $-40^\circ$C 
has dried out with 170~W dummy load power, which is
significantly less than the expected cooling power
of 280~W (=200~J/g$\times$1.4~g/s).
The reason of this deficit of the cooling power
is presumably heat load of the transfer tube
and other low temperature part. 
This measurement was done at the ambient temperature 
of $27^\circ$C, while the temperature of the liquid
CO$_2$ was $15^\circ$C. 
%\begin{wrapfigure}{r}{0.45\columnwidth}

Pressure drop in the metal core flexible tubes 
was one of concerns because inner wall of the tubes
has corrugated structure. The pressure drop has
actually measured for several values of the flow rate.
The measured pressure drop is reasonably low. 
At low flow rate ($< 1.4$~g/s), the pressure drop
is less than resolution of the digital pressure gauge.
At high flow rate ($> 2$~g/s), the pressure drop
is dominated by the flow meter for liquid, 
and still less than the resolution of the pressure gauge
for gas.
 
\section{Further R\&D}
\subsection{Pressure control}
In the prototype cooling system, pressure of
the two-phase CO$_2$ is manually controlled by a 
back-pressure valve. This method is however somewhat
unstable and time consuming for adjustment.
To overcome this disadvantage, we plan to
replace the manual back-pressure valve with
an automatic pressure controller.
We have tested a commercially available pressure
controller, Bronkhorst P-702CV. Using this controller,
the back pressure can be controlled by external
voltage setting.

\begin{figure}[t]
\centerline{\includegraphics[width=0.77\columnwidth]{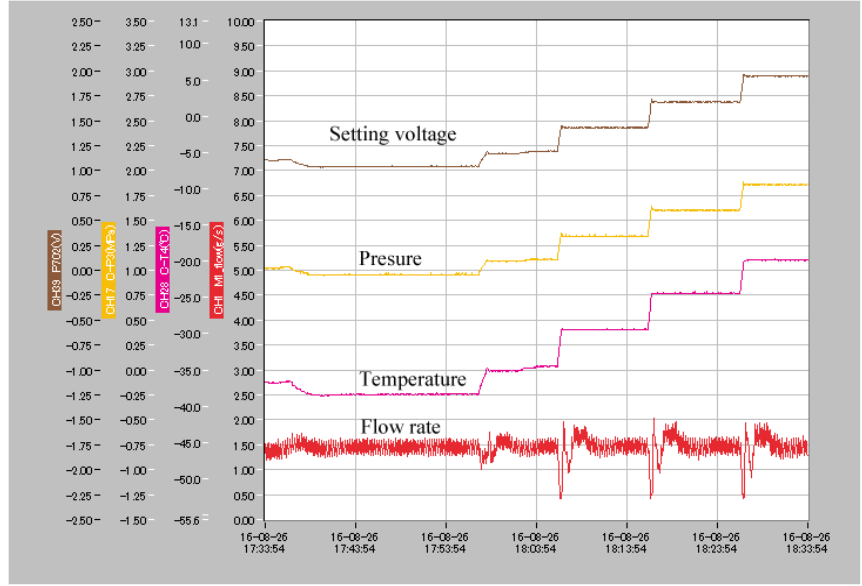}}
\caption{Setting voltage for the pressure controller, 
and the response of pressure and temperature of two-phase CO$_2$
to the setting voltage. The flow rate is also shown.}
\label{fig:pcontrol}
%\end{wrapfigure}
\end{figure}

We have studied this pressure controller 
by inserting it in the return gas line between
the cooling unit and the liquefier unit. 
The back-pressure valve in the liquefier unit
was set at the lowest pressure.
Figure~\ref{fig:pcontrol} shows the result of the
measurement of the setting voltage, pressure 
and temperature of the two-phase CO$_2$, 
and flow rate of CO$_2$.
It can be seen that quick and stable control
of the pressure and the temperature is
achieved.
%\begin{wrapfigure}{r}{0.5\columnwidth}

\subsection{O-ring material}
O-rings made of elastomer are used for the gas compressor
and safety valves in our system. 
Degradation of the O-rings called as explosive decompression (ED) 
was seen in these O-rings, and caused gas leak.
Explosive decompression, also called as rapid gas decompression,
is a mechanism of degradation in elastomer due to rapid
decompression of gaseous media. 
At high pressure environment, CO$_2$ gas immigrates into
elastomer. When the pressure is reduced suddenly, 
CO$_2$ dissolved inside the elastomer comes out
as micro bubbles, expands, and damage the
elastomer from inside.

In order to mitigate the risk of gas leak due to ED,
we can replace the safety valves with metal-seal
safety valves. For the gas compressor, we should
find out better material for O-rings.
We have constructed O-ring ED test apparatus,
and plan to test several kinds of O-rings with
different Shore durometer hardness and
different materials such as Kalrez, Chemraz, 
and so on.
In the worst case, frequent overhaul of the gas compressor 
would be the solution.

\subsection{Other R\&D issues}
In the present prototype system, a very massive stainless-steel plate
heat exchanger is used. We plan to develop a very low mass
heat exchanger which can be placed inside the detector.
As a candidate, a heat exchanger made of double-layer tube
will be studied.

The size of the liquefier unit for the present prototype
is quite large ($\sim$1~m$\times$1~m$\times$2~m). 
We would like to develop a more compact liquefier unit.
The gas booster used in the liquefier is quite noisy.
Sound insulation should be implemented in the next prototype
of the liquefier unit.

The present prototype is  controlled manually.
The remote control system using a programmable logic
controller (PLC) is one of the R\&D issues.

On the detector side, thermal contact between the cooling tube
and the end-plate of the vertex detector, and between
the end-plate and the ladders has to be studied.
 
\section{Summary}
We have successfully developed a prototype of
two-phase CO$_2$ cooling system using a
gas compressor for CO$_2$ circulation.
Cooling power of the system has been measured 
at the cooling temperature of $-40^\circ$C,
and a satisfactory results have been obtained
for FPCCD vertex detector cooling.
On the other hand, degradation problem of
O-rings exists, and has to be solved.
There are still many R\&D issues to be
accomplished to realize the practical
cooling system.
 
\section*{Acknowledgments}
This work is  supported by Grant-in-Aid No.24540312
and No.16H03992 by Japan Society for 
Promotion of Science (JSPS). 
This work is also  supported by KEK 
Detector Technology Project.
 
% ****************************************************************************
% BIBLIOGRAPHY AREA
% ****************************************************************************
\begin{footnotesize}
% IF YOU DO NOT USE BIBTEX, USE THE FOLLOWING SAMPLE SCHEME FOR THE REFERENCES
% ----------------------------------------------------------------------------

% ----------------------------------------------------------------------------
\end{footnotesize}

% ****************************************************************************
% END OF BIBLIOGRAPHY AREA
% ****************************************************************************

\begin{thebibliography}{99}
% Please replace the numbers for   contribId   and   sessionId
% in the following URL. You can get this information by going to 
% http://indico.cern.ch/confAuthorIndex.py?confId=2628
% and search for your contribution and click on the title
% Be aware: '&amp;' must be replaced by simple '&' as in example below
%------- replace following references ;-)
\bibitem{behnke} T.~Behnke (ed.) \textit{et~al.}, The International 
Linear Collider Technical Design Report-Volume~4:Detectors,
arXiv:1306.6329[physics.ins-det] (2013). 
\bibitem{sugimoto05} Y.~Sugimoto, Proceedings of International 
Linear Collider Workshop LCWS05, Stanford, CA, March 2005, pp.550-554.
\bibitem{attree} D.~Attree \textit{et~al.}, 2008 JINST 3 P07003.
\bibitem{delli} A.A.M.~Delli \textit{et~al.}, NLR-TP-2003-001 (2003).
\bibitem{beuzekom} M.~Van Beuzekom \textit{et~al.}, Pos (Vertex2007) 009 (2007). 
\bibitem{sugimoto12} Y.~Sugimoto \textit{et~al.}, arXiv:1202.5832[physics.inst-det] (2012).
\bibitem{sugimoto12b} Y.~Sugimoto, CO$_2$ cooling for FPCCD Vertex Detector, talk given at KILC12 Workshop, April 24th, 2012.
\bibitem{sugimoto13} Y.~Sugimoto, R\&D status of FPCCD VTX and its cooling system, talk given at ECFA LC2013 Workshop, May 28th 2013.
\end{thebibliography}
\end{document}